\documentclass[]{elsart}
\usepackage{amssymb}
\usepackage{pst-node,pstricks,graphicx}

\begin{document}                
\begin{frontmatter}
\title{Primary Particle Type of the Most \\ Energetic Fly's Eye Air Shower}

\author[a,b]{M.~Risse,\corauthref{cor1}}
\author[b]{P.~Homola,}
\author[b]{D.~G\'ora,}
\author[b]{J.~P\c{e}kala,}
\author[b]{B.~Wilczy\'nska,}
\author{and}
\author[b]{H.~Wilczy\'nski} 
\corauth[cor1]{ {\it Correspondence to}: M.~Risse
(markus.risse@ik.fzk.de)}
\address[a]{
Forschungszentrum Karlsruhe, Institut f\"ur Kernphysik, 76021 Karlsruhe,
Germany
}

\address[b]{
H.~Niewodnicza\'nski Institute of Nuclear Physics, Polish Academy of Sciences,
ul.~Radzikowskiego 152,
31-342 Krak\'ow, Poland
}

\begin{abstract}                
The longitudinal profile of the most energetic cosmic-ray air shower
measured so far, the event recorded by the Fly's Eye detector with a
reconstructed primary energy of about $3.2\cdot10^{20}$~eV,
is compared to simulated shower profiles. 
The calculations are performed with the CORSIKA code
and include primary photons and different hadron primaries. 
For primary photons, preshower formation in the
geomagnetic field is additionally treated in detail.
For primary hadrons, the hadronic interaction models
QGSJET~01 and SIBYLL~2.1 have been employed.
The predicted longitudinal profiles are compared to the observation.
A method for testing the hypothesis of a specific primary particle
type against the measured profile is described
which naturally takes shower fluctuations into account.
The Fly's Eye event is compatible with any assumption of a hadron primary
between proton and iron nuclei in both interaction models,
although differences between QGSJET~01 and SIBYLL~2.1 in the predicted profiles
of lighter nuclei exist.
The primary photon profiles differ from the data on a level of
$\simeq$ 1.5$\sigma$.
Although not favoured by the observation, the primary photon hypothesis
can not be rejected for this particular event.
\end{abstract}
\end{frontmatter}
\section{Introduction}
\label{sec-intro}
Identifying the primary particle type might provide a key to understanding
the origin of the extreme high-energy cosmic rays (EHECR) with energies 
around and exceeding $10^{20}$~eV.
In ``bottom-up'' acceleration scenarios, hadrons are the favoured
particles at production site with different predictions ranging from
proton to iron-dominated fluxes.
On the contrary, in ``top-down'' scenarios generally a large fraction of
the observable events are predicted to be photon-initiated.
In these scenarios, a decay of supermassive ``X-particles'' is assumed
which supposedly were created by topological defects such as cosmic strings and
magnetic monopoles or produced as metastable particles in the early Universe.
In the decay chain of the ``X-particles'', a large fraction of
EHE photons is predicted.
For a review, see e.g.~\cite{topdown} and references given therein.
It should be noted that also in ``bottom-up'' scenarios a small
fraction of the particles entering the atmosphere might
be photons due to interaction processes 
during the hadron propagation. Also ``top-down'' scenarios with a reduced
fraction of predicted EHE photons exist.
However, the determination of the nature of the EHECR and 
especially conclusions on the fraction of EHE photons will provide
strong constraints on models of cosmic-ray origin and help to decide
between ``top-down'' and ``bottom-up'' scenarios.

Based on an analysis of muons in air showers observed by the
Akeno Giant Air Shower Array (AGASA), upper limits on the photon flux
were estimated to be 28\% above $10^{19}$~eV and 67\% above $10^{19.5}$~eV
at a confidence level of 95\%~\cite{agasa}.
By comparing the rates of near-vertical showers to inclined ones
recorded by the Haverah Park shower detector, upper limits of
48\% above $10^{19}$~eV and 50\% above $4\cdot10^{19}$~eV
(95\% C.L.) were deduced~\cite{havpark}.

The most energetic cosmic ray reported so far was measured 
by the Fly's Eye detector~\cite{detector,event} located at
Dugway, Utah (40$^{\circ}$~N, 113$^{\circ}$~W).
The observation was performed with a compound eye of
880 photomultiplier tubes monitoring the sky.
The tubes were arranged at the focal planes of
67 mirrors, each mirror having 1.5~m diameter.
An air shower is observed by the Fly's Eye
as fluorescent light source
where the local light intensity is closely connected to the charged
particle content of the shower, thus allowing to derive a longitudinal 
shower profile.

The record event was detected on October 15, 1991, at 7:34:16~UT
with a reconstructed energy of about $(3.2\pm0.9)\cdot10^{20}$~eV~\cite{event}.
The main characteristics of this event are summarized in Table~\ref{tab-event}.
Comparing the reconstructed depth of shower maximum with expectations
from air shower simulations, it has been concluded
that the event agrees well with hadron-initiated showers. 
The best fits were reached for mid-size nuclei, but due to
the experimental uncertainties and due to shower fluctuations, even
nucleons or heavy nuclei were not excluded \cite{event}.
Similar conclusions were obtained in \cite{halzen}.
For the hypothesis of a primary photon, it was pointed out
that the event direction is nearly perpendicular to the local
geomagnetic field \cite{event,halzen,karakula}.
Thus, the possibility of creating an
electromagnetic cascade (``preshower'') in the magnetosphere before
entering the atmosphere 
has to be taken into account, with considerable effects on the
longitudinal shower profile. In a more quantitative estimate, 
the predicted cascade curves for primary photons were claimed to be in
disagreement with the measurements~\cite{halzen}.

The preshower calculation in~\cite{halzen} was performed based
on~\cite{erber}.
Recently, a numerical error in~\cite{erber} has been noted that affects
the evaluation of photon conversion probability~\cite{preshcors}.
An independent cross-check of the conclusion of~\cite{halzen},
that are widely referred to in the literature 
(see, e.g., ~\cite{topdown,watson03})
seems desirable.
\begin{table}[t]
\begin{center}
\caption{The main shower parameters of the most energetic Fly's Eye
event~\cite{event}. The combined uncertainty is obtained adding the
statistical and systematic uncertainties in quadrature. 
The azimuth angle is given counterclockwise from east for the incoming
direction.}
\label{tab-event}
\vskip 0.5 cm
\begin{tabular}{ccccc}
\hline
          &  Best-Fit & Statistical & Systematic & Combined
\\
Parameter &  Value    & Uncertainty & Uncertainty& Uncertainty
\\
\hline\hline
Energy [$10^{18}$~eV]& 320 & $^{+35}_{-40}$& $\pm 85$ &$^{+92}_{-94}$
\\
$X_{max}$ [g/cm$^2$] & 815 & $^{+45}_{-35}$& $\pm 40$ &$^{+60}_{-53}$
\\
zenith angle [deg] & 43.9 & $^{+1.4}_{-0.6}$& $\pm 1.2$ &$^{+1.8}_{-1.3}$
\\
azimuth angle [deg]& 31.7 & $^{+4.0}_{-6.0}$& $\pm 1.2$ &$^{+4.2}_{-6.1}$
\\
\hline
\\
\end{tabular}
\end{center}
\end{table}

In this work, the reconstructed longitudinal profile of the Fly's Eye event
is analyzed employing the CORSIKA shower simulation code~\cite{corsika},
which is linked to the PRESHOWER program~\cite{preshcors} in case of primary
photons.
The fact already expressed
in~\cite{event} that the shower maximum depth does not uniquely identify the
type of the primary particle, can not be overcome. However, an approach
is described here to give an approximate probability of the measured event
to be consistent with the expectation of a specific primary particle type.
This approach will naturally take air shower fluctuations into account,
which in the hadronic case are known to be larger for the lighter nuclei.
Two different hadronic interaction models, QGSJET~01~\cite{qgsjet01} and
SIBYLL~2.1~\cite{sibyll2.1}, are used in the comparison.
Special attention is paid to the investigation of the primary photon
hypothesis. 

The plan of the paper is as follows.
The calculation of the longitudinal shower profiles is described
in Section~\ref{sec-calculation}. This includes a brief description
of the preshower physics and of the CORSIKA code.
The results are given in Section~\ref{sec-results}, both for hadrons
using different primary masses and interaction models and for primary
photons.
The depth of shower maximum is investigated in Section~\ref{subsec-xmax}.
Based on the complete longitudinal profile, 
the level of \mbox{(dis-)}agree\-ment between simulations and observation
is tentatively quantified in Section~\ref{subsec-profile}.
Conclusions are discussed in Section~\ref{sec-conclusion}.

It should be pointed out that the current paper investigates the 
possibility of assigning a particle type to the
event profile published in Ref.~\cite{event},
taking the reconstruction results at face value.
No attempt is made to analyze the reconstruction method itself;
for instance, no conclusion about the reconstructed primary energy
is made.
Also we do not comment on potential astrophysical sources
of this event as well as particle dependent propagation effects on the way
from the source to the observer; this has been discussed in detail
elsewhere~\cite{sommers95}.

\section{Calculation of longitudinal profiles}
\label{sec-calculation}
\subsection{Preshower formation}
\label{subsec-preshower}
At energies above $10^{19}$~eV
in the presence of the geomagnetic field, a photon can convert into an 
electron-positron pair before entering the atmosphere. 
For a small path length $dr$, the conversion probability
can be determined as~\cite{erber}
\begin{equation}
\label{pconv}
p_{conv}(r)
\simeq\alpha(\chi(r))dr
\end{equation}
where $\alpha(\chi) = 0.5(\alpha_{em}m_ec/\hbar)(B_\bot/B_{cr})T(\chi)$, 
$\chi\equiv0.5(h\nu/m_ec^2)(B_\bot/B_{cr})$, $\alpha_{em}$ is the fine structure 
constant, $B_\bot$ is the magnetic field component transverse to the
direction of photon motion, $B_{cr} \equiv m_e^2c^3/e\hbar=4.414 \cdot 10^{13}$ G, 
and $T(\chi)$ is the magnetic pair production function which is negligible if
$\chi \ll 1$, has a maximum around $\chi=5$ and then decreases slowly to zero 
\cite{erber,preshcors}.
The resultant electrons will 
subsequently lose their energy by magnetic bremsstrahlung (synchrotron radiation). 
The probability of emitting a bremsstrahlung photon 
by a single electron over a small distance $dr$ is given by \cite{sokolov}:
\begin{equation}
\label{pbrem}
p_{brem}(B_\bot,E,h\nu,dr)=dr \int^E_0 I(B_\bot,E,h\nu)\frac{d(h\nu)}{h\nu}
\end{equation}
where $I(B_\bot,E,h\nu)$ is the spectral distribution of radiated energy, $E$ is
the electron energy and $h\nu$ is the energy of the emitted bremsstrahlung
photon \cite{sokolov,preshcors}.
If the energy of the emitted photon is high enough, it can create
another electron-positron pair. In this way, instead of
the primary high-energy photon, a number of less energetic 
particles, mainly photons and a few electrons, will enter the atmosphere. We
call this cascade a ``preshower'' since it originates and develops above the
atmosphere, i.e.~before the ``ordinary'' shower development in air.

\begin{figure}[t]
\begin{center}
\includegraphics[height=8.7cm,angle=0]{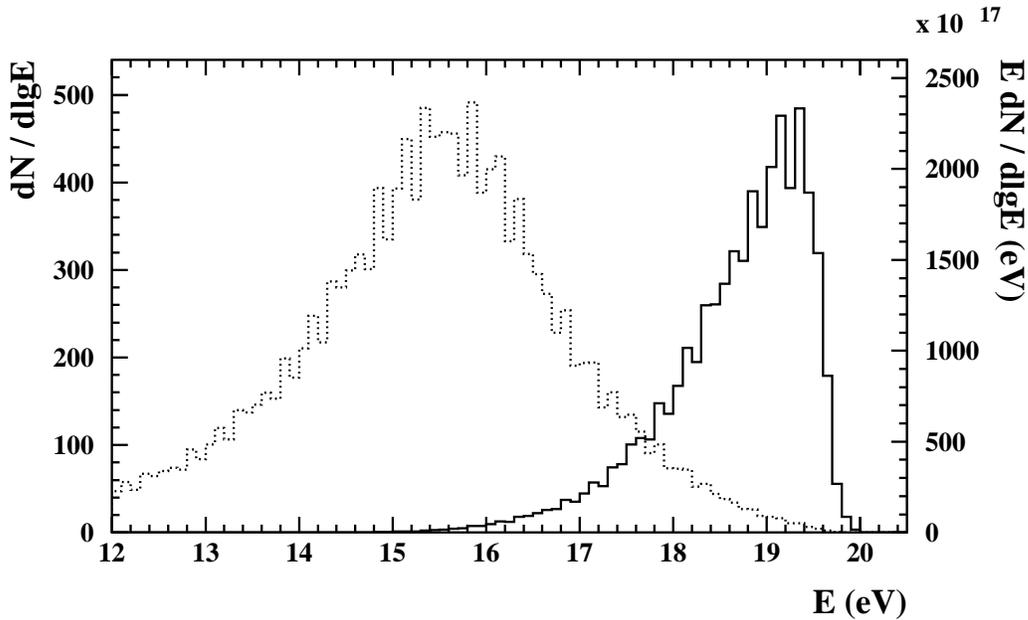}
\caption{Energy spectrum of the preshower particles (dotted line) and
spectrum weighted by energy (solid line) for the conditions of the
Fly's Eye event (average of 1000 simulation runs).}
\label{fig-espek}
\end{center}
\end{figure}

Preshower features have been investigated by various 
authors, see for instance
\cite{halzen,karakula,preshcors,mcbr,ahar,van1,stan,bert,plya,bedn,van2}.
A detailed description of the code applied in the present analysis
is given in~\cite{preshcors}.
In brief, the geomagnetic field components are calculated according to
the International Geomagnetic Reference Field (IGRF) model~\cite{nasa}.
The primary photon propagation is started at an altitude corresponding
to a preshower path length of five Earth's radii.
The integrated conversion probability at larger distances is sufficiently small.
The photon conversion probability is
calculated according to Eq.~(\ref{pconv}).
For the Fly's Eye event conditions, all simulated primary photons converted.
After conversion, the resultant electrons 
are checked
 for bremsstrahlung emission with a
probability distribution of the emitted photon energy following Eq.~(\ref{pbrem}).
A cutoff of 10$^{12}$~eV is applied for the bremsstrahlung photons,
as the influence of photons at lower energies is negligible
for the air shower evolution.
The preshower simulation is finished when the top of atmosphere is reached. 
Then, all preshower particles are passed to CORSIKA. 
The resultant energy spectrum of the preshower particles is shown in
Figure~\ref{fig-espek}. The photons and electrons reach the atmosphere
with energies below $10^{20}$~eV; most of the initial energy is stored
in particles of $\simeq 10^{19}$~eV.

\subsection{Event generation with CORSIKA}
\label{subsec-corsika}
CORSIKA, a standard tool for Monte Carlo shower simulation, has been
used for calculating the atmospheric shower profiles (version 6.16).
In case of primary photons, the resultant atmospheric cascade is
simulated as a superposition of subshowers initiated by the preshower particles.
Electromagnetic interactions are treated by the EGS4 code~\cite{egs}, which
has been upgraded to take photonuclear reactions
as well as the Landau-Pomeranchuk-Migdal (LPM) effect~\cite{lpm} into 
account~\cite{corsika,lpmcorsika}.
The LPM effect leads to an increase of the mean free path of electromagnetic
particles and enhances an asymmetric energy distribution of the secondary particles.
In particular for electromagnetic particles at energies exceeding $10^{19}$~eV,
the shower development can be considerably retarded, and shower-to-shower
fluctuations are large. For the primary energy and direction of the Fly's Eye event,
however, the
preshower formation in case of primary photons transfers the energy to
particles which are less (or not at all) affected by the LPM effect.

\begin{figure}[t]
\begin{center}
\includegraphics[height=9.7cm,angle=0]{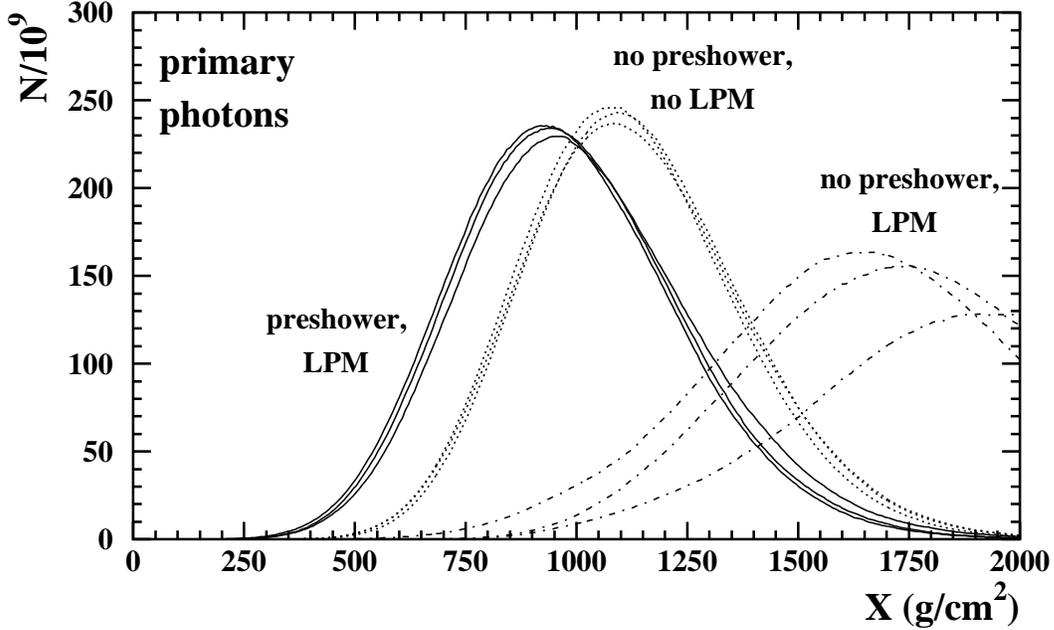}
\caption{Examples of longitudinal shower profiles of primary photons
for the Fly's Eye event conditions with and without simulating 
preshower formation in the geomagnetic field.}
\label{fig-wwopre}
\end{center}
\end{figure}

To illustrate the importance of accounting for the preshower formation,
longitudinal profiles of primary photons without preshower simulation are
displayed in Figure~\ref{fig-wwopre} together with results of the complete
calculation.
It is interesting to note that the depth of shower maximum in the full
simulation is significantly smaller also compared to the case of 
neglecting both preshower and LPM effect: Most shower energy is stored
in particle energies 1.5 orders of magnitude below the initial one (see
Figure~\ref{fig-espek}). Thus, with an elongation rate of
$\simeq$85~g/cm$^2$ per primary energy decade for unaffected photon showers,
a rough estimate yields differences of $\simeq$130~g/cm$^2$
in depth of shower maximum, in good agreement with the simulation.

Hadron-initiated showers were simulated with the hadronic interaction
models QGSJET~01~\cite{qgsjet01} and SIBYLL~2.1~\cite{sibyll2.1}
to study the difference in primary type assignment due to different
modeling of high-energy hadronic interactions.
Simulation thresholds for the kinetic particle energies
of 250~keV for electromagnetic particles and
100~MeV for hadrons and muons were adopted.
The {\it thin sampling} option was used which allows Monte Carlo
profile calculations at highest energies within acceptable computing times.
In this option, the bulk of secondary particles produced in an interaction is
discarded, and only ``representative'' particles are sampled.
Weight factors are assigned to the selected particles to keep energy
conservation~\cite{hillas}.
Thin sampling is activated, however, only at particle energies much smaller
than the primary energy. This ensures an undisturbed modeling of
the shower profile fluctuations, in particular those at the initial states
of the shower cascade. Additionally, upper limits for the particle weights
have been chosen in CORSIKA to keep the contribution to the profile of
an individual particle sufficiently small~\cite{kobal,risseicrc01,gap-edep}.
More specifically, thinning is activated for particles with energies
$< 3\cdot10^{-6}E_0$, with $E_0$ being the primary energy, and
particle weights are limited to $<10^{6}$, which is negligible compared to the
total number of particles in the shower development range relevant for the
current analysis.

Based on the reconstructed values of primary energy and direction
quoted in Table~\ref{tab-event},
simulations were performed for primary protons, carbon
and iron nuclei, both for QGSJET~01 and SIBYLL~2.1,
and for primary photons (using QGSJET~01 for the small amount
of hadron-hadron interactions). For each primary particle setting,
1000~events were generated. Additionally, primary energy
and direction were varied within the given reconstruction
errors. For all primary particle types it turned out that the
calculated profiles showed no significant changes, apart from
the normalization in case of primary energy variations,
within the parameter range allowed by the reconstruction uncertainties.
This is expected for hadron-initiated events.
Given an elongation rate of about 50$-$60~g/cm$^2$ per primary energy
decade, a 10\% change in energy corresponds to a shift of
shower maximum of $\simeq$2~g/cm$^2$,
which is well within the measurement uncertainties and intrinsic
shower fluctuations.
Primary photon profiles are very sensitive to variations of the
primary energy or arrival direction close to
the threshold for preshower formation. The Fly's Eye event conditions,
however, are well above the preshower onset also when varying the
primary energy and direction within the reconstruction uncertainties.

\section{Comparison of measurement and simulation}
\label{sec-results}
\subsection{Depth of shower maximum $X_{max}$}
\label{subsec-xmax}

\begin{figure}[t]
\begin{center}
\includegraphics[height=10cm,angle=0]{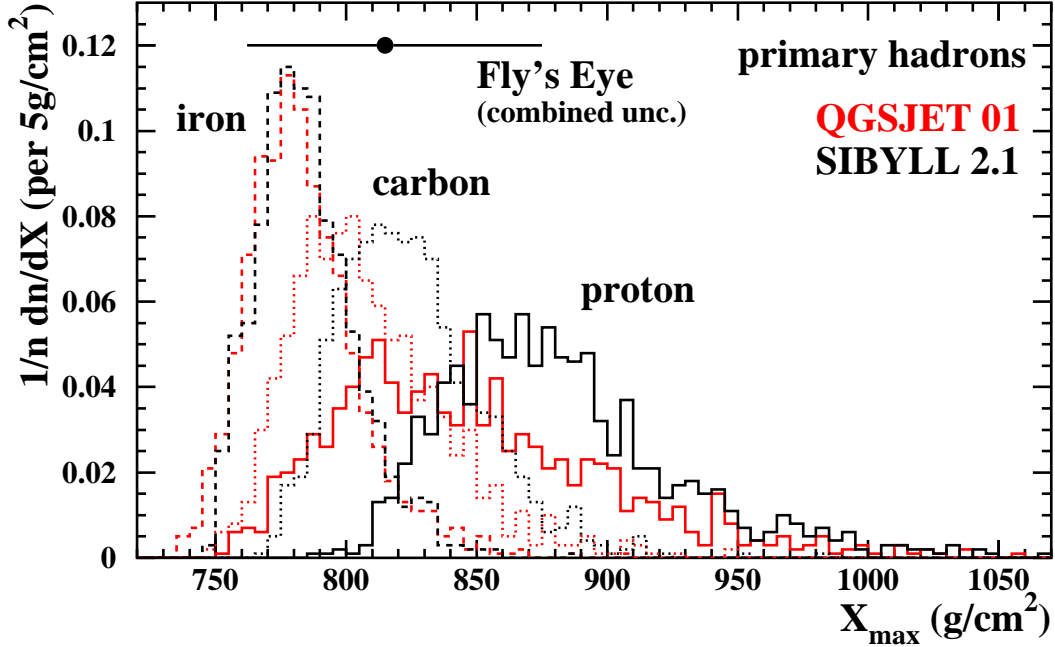}
\caption{Shower maximum distributions of primary hadrons (iron $-$ dashed, 
carbon $-$ dotted, proton $-$ solid line), both for QGSJET~01 and SIBYLL~2.1.
The measured value is indicated with the combined uncertainty,
cf.~Table~\ref{tab-event}.}
\label{fig-xmaxha}
\end{center}
\end{figure}

The distributions of depth of shower maximum for various nuclear primaries
and for the hadronic interaction
models QGSJET~01 and SIBYLL~2.1 are shown in Figure~\ref{fig-xmaxha}.
The reconstructed depth of the Fly's Eye event of 815~g/cm$^2$ corresponds to a predicted average value
of a mid-size nucleus.
But due to the measurement uncertainties and due to the intrinsic shower fluctuations,
any of the hadron primaries considered in both models could have produced
the observed shower depth.

The average depth of shower maximum and its fluctuations are listed in
Table~\ref{tab-xmax}.
While the predictions for heavy nuclei agree for the two models, the difference
for primary proton amounts to about 35~g/cm$^2$. Compared to SIBYLL~2.1,
composition analyses based on QGSJET~01 
would deduce somewhat lighter nuclei as primary particles
using average shower maximum depths.
However, for the current investigation this is of minor importance.

\begin{table}[t]
\begin{center}
\caption{Average depth and RMS of shower maximum for the given
primary particle assumptions and discrepancy $\Delta$ to the observed shower
maximum in units of standard deviations.}
\label{tab-xmax}
\vskip 0.5 cm
\begin{tabular}{lccc}
\hline
&        &  ~~~QGSJET~01~~~  &  ~~~SIBYLL~2.1~~~ \\
& photon & p~~~~~C~~~~~Fe & p~~~~~C~~~~~Fe \\
$X_{max}$ [g/cm$^2$] & 937 & 848~~~808~~~783
         & 882~~~824~~~785
\\
RMS($X_{max}$) [g/cm$^2$] & 26 & 54~~~~~30~~~~~22
         & 47~~~~~27~~~~~19
\\
$\Delta (X_{max})$ [$\sigma$] & 1.9 & 0.6~~~~0.4~~~~0.6 & 0.9~~~~0.4~~~~0.6
\\
\hline
\\
\end{tabular}
\end{center}
\end{table}
\begin{figure}[t]
\begin{center}
\includegraphics[height=9.6cm,angle=0]{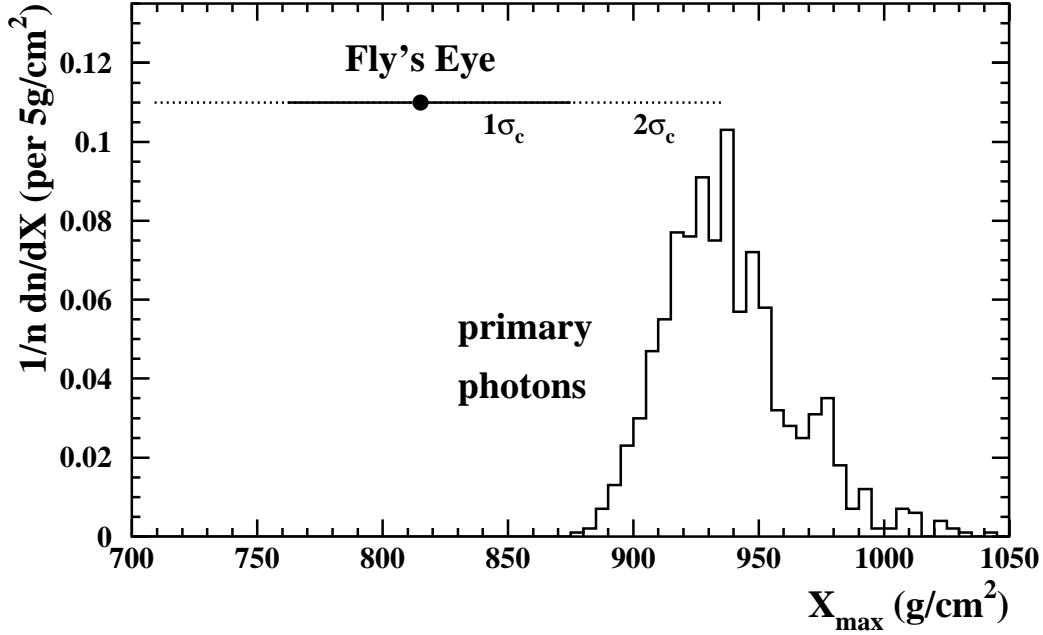}
\caption{Shower maximum distribution of primary photons
compared to the reconstructed value of the Fly's Eye event.
The measured depth is shown with the combined uncertainty
of $1\sigma_c$ and $2\sigma_c$,
cf.~Table~\ref{tab-event}.}
\label{fig-xmaxph}
\end{center}
\end{figure}

In Figure~\ref{fig-xmaxph}, the $X_{max}$ distribution of primary photons
is displayed. The Fly's Eye event depth of maximum of 815~g/cm$^2$
is smaller than the average $X_{max}$ of primary photons of 937~g/cm$^2$.
Given a combined uncertainty of 60~g/cm$^2$ for the measured $X_{max}$,
the difference between these values is about two standard deviations.

To account also for shower fluctuations, the following
method is applied to quantify the level of agreement between
a primary particle hypothesis and the observation.
For each profile $i$ with its individual shower maximum,
the corresponding probability $P_i$ of consistency with data is determined.
The total probability $P$ upon which the decision is based whether to reject
a specific primary particle hypothesis is then calculated from
all $n_s = 1000$ profiles simulated per primary type as
\begin{equation}
\label{eq-ptot}
P = \frac{1}{n_s} \sum_{i=1}^{n_s} P_i ~.
\end{equation}
In this approach, also possible non-Gaussian shower fluctuations predicted by
the simulations are naturally taken into account.
An incorrect consideration of shower fluctuations might introduce a significant
bias in the primary type assignment, since the fluctuations are different
for the various primary particles.

For primary photons, adopting the combined uncertainty of the data
given in Table~\ref{tab-event}, a value of $P_{X_{max}} \simeq 6\%$
is obtained.
This corresponds to a discrepancy of $\Delta \simeq 1.9\sigma$, where
$\Delta$ is defined by $P(\chi^2 \ge \Delta^2) = 6\%$ from the $\chi^2$
distribution for one degree of freedom.
The discrepancy between data and calculation is listed in Table~\ref{tab-xmax}
for the different primary particle assumptions.

It might be instructive to illustrate the importance of reducing the
measurement uncertainty.
Assuming a resolution in $X_{max}$ of $\le 30$~g/cm$^2$ as envisaged
for the detectors of the Pierre Auger Observatory~\cite{auger,augerea},
the discrepancy of primary photons to the reconstructed value
would amount to $\ge 3.3\sigma$ instead of $1.9\sigma$.

\subsection{Complete longitudinal profile}
\label{subsec-profile}
\begin{figure}[t]
\begin{center}
\includegraphics[height=10cm,angle=0]{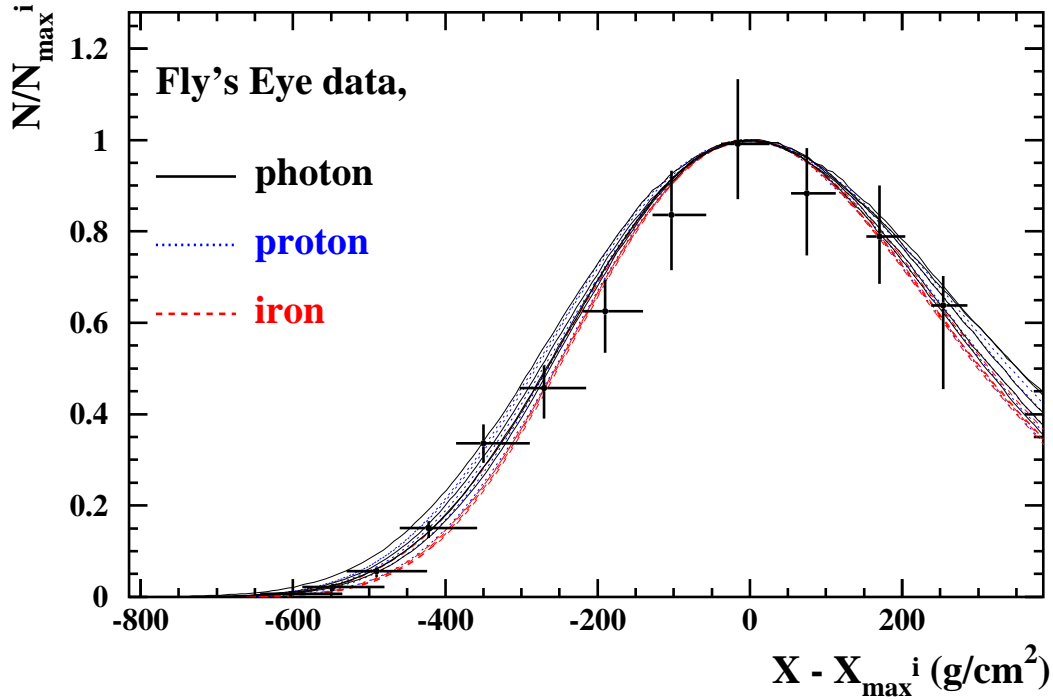}
\caption{Profile shape of simulated showers initiated by
photon, proton, and iron primaries (calculated with QGSJET~01)
in comparison to the data.
Each profile has been divided by the individual maximum number of 
particles $N_{max}^i$ and has been shifted by the respective
depth of shower maximum $X_{max}^i$.}
\label{fig-shape}
\end{center}
\end{figure}

For the various hadron primaries, differences in the average number
of particles (shower size) at maximum are known to be on the level
of only a few percent and thus much smaller than the uncertainty
in primary energy reconstruction of the Fly's Eye event.
In case of unconverted photons, the number of particles at maximum
could be considerably reduced, which is accompanied by a change in the shape
of the profile, as can be seen in Figure~\ref{fig-wwopre}.
For the specific conditions of the Fly's Eye event, however, all simulated
photons converted, and the differences in average shower
size at maximum compared to primary hadrons are small ($<$8\%).
Also the profile shapes of the different primaries are in reasonable
agreement to the data, as shown in Figure~\ref{fig-shape}.
Iron-initiated showers show slightly narrower profiles than, e.g., 
proton-induced ones. The difference is small, however, compared to the
measurement uncertainties.

It seems worthwhile to mention that a quantity such as
``number of particles'' would require a precise definition 
if conclusions were attempted based on a direct comparison
of shower sizes provided by different approaches.
For instance, the number of tracked particles in Monte Carlo
calculations depends on the simulation energy threshold~\cite{nerling}.
Also, shower particles in three-dimensional calculations show
an angular spread around the shower axis~\cite{muniz,erelease},
which should be kept in mind when comparing to an ``effective''
particle number in a one-dimensional calculation.
However, due to the relatively large primary energy uncertainty of
Fly's Eye event, these items are of minor relevance for the present
analysis.

\begin{figure}[t]
\begin{center}
\includegraphics[height=15cm,angle=0]{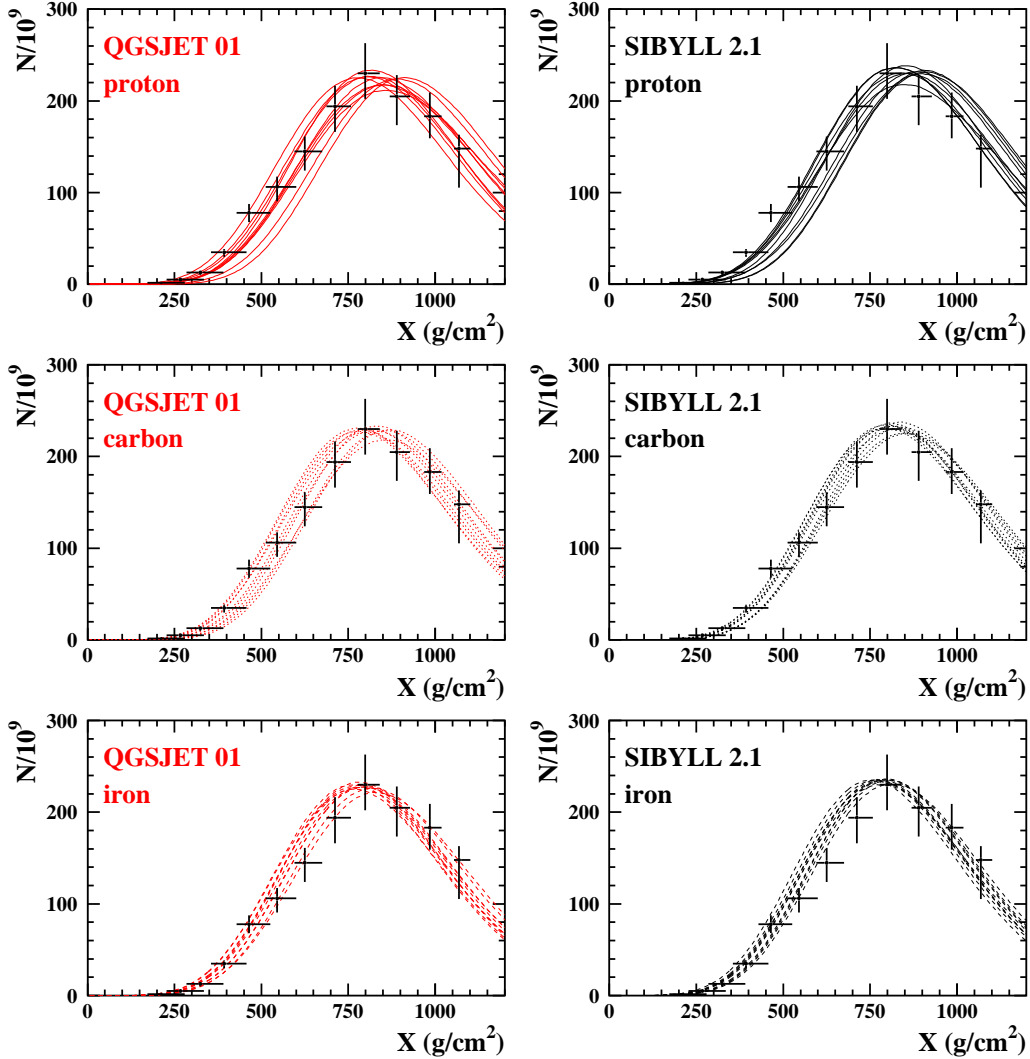}
\caption{Typical longitudinal profiles of hadron-initiated showers
compared to the data for different combinations of primary particle
and hadronic interaction model as assigned.} 
\label{fig-profileha}
\end{center}
\end{figure}

\begin{figure}[t]
\begin{center}
\includegraphics[height=10cm,angle=0]{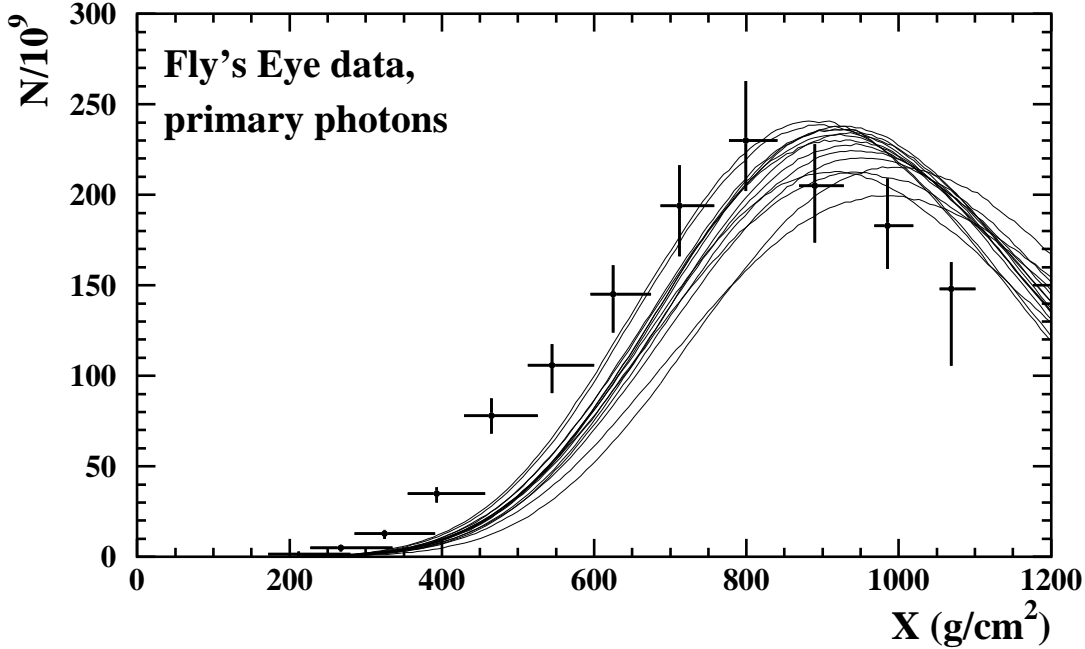}
\caption{Typical longitudinal profiles of photon-initiated showers
compared to the data.} 
\label{fig-profileph}
\end{center}
\end{figure}

The complete longitudinal event profile is displayed together with
simulated hadron-induced showers
in Figure~\ref{fig-profileha} and with photon-initiated events in
Figure~\ref{fig-profileph}. 
Only a subset of the simulated events, picked at random, is plotted.
The normalization of the calculated profiles has been chosen such that
the number of particles at shower maximum roughly coincides with the
reconstructed one of the Fly's Eye event. More specifically,
all CORSIKA shower size profiles are scaled by a factor 1.06 in case
of primary photons and by 1.10 for primary hadrons,
which is well within the reconstruction uncertainty of the primary energy.


The profiles of the different hadron primaries show reasonable agreement
with the data for both interaction models.
In case of primary photons (Figure~\ref{fig-profileph}),
differences between data and simulation are visible
that can be attributed to a shift in atmospheric depth
of the whole profile. Since the measured data points are
constrained by a common geometry fit, it is important to note that their
reconstructed atmospheric depth values are correlated~\cite{event}.
Qualitatively, a shift in depth of each point by about 1.5 units of its
standard deviation would results in a reasonable
agreement of data and the photon profiles.
In the following, a quantitative treatment for testing
a primary particle assumption against the observed profile
is described.

Shower fluctuations are taken into account as discussed in 
Section~\ref{subsec-xmax} by averaging the probabilities $P_i$ for
the individual profiles to obtain an overall probability $P$
for each considered primary particle type (Eq.~\ref{eq-ptot}).
The calculation of $P_i$ for each individual simulated profile is based
on the reconstructed profile data as shown in Figure~\ref{fig-profileph}
assuming Gaussian probability density functions for the quoted uncertainties
$dN_j$ and $dX_j$ of the data points.
For the uncertainties $dX_j$, we take the point-to-point correlation into
account and allow all data points to be shifted in the
same direction by the same fraction $f\cdot dX_j$ when varying
the reconstructed geometry.
Given a simulated profile $i$, for a certain correlated shift of
the $n_d = 12$ data points in horizontal direction by
$f_k \cdot dX_j$, a value $\chi^2_{ik}$ is then calculated as
\begin{equation}
\label{eq-chi2}
\chi^2_{ik} = \sum_{j=1}^{n_d}\left(\frac{\Delta N_{jk}}{dN_j}\right)^2
\end{equation}
with $\Delta N_{jk} = N_j - N_{sim}(X_j+f_k \cdot dX_j)$ denoting
the difference between the reconstructed event value $N_j$ and the
simulated value $N_{sim}$ at depth \mbox{$X_j+f_k \cdot dX_j$}.
Factors $f_k = -3.0,...,+3.0$ are adopted with a stepsize of
$\Delta f = f_{k+1} - f_k = 0.05$.
From $\chi^2_{ik}$, the probability $P_{ik}$ that the $i$-th profile fits
the shifted data is determined as
\begin{equation}
\label{eq-pik}
P_{ik} = P(\chi^2_{ik,{\rm gen}} \ge \chi^2_{ik})
\end{equation}
by randomly generating a series of artificial data sets according to
the uncertainties $dN_j$ around $N_{sim}(X_j)$ and shifting each data set 
in the same way as the real data to obtain the distribution
of $\chi^2_{ik,{\rm gen}}$ values~\cite{lyons}.
The probability $P_i$ is then given by the sum
\begin{equation}
\label{eq-pi}
P_i = \sum_{k} w_k \cdot P_{ik}
\end{equation}
where the individual $P_{ik}$ are weighted for step $k$ according to
a Gaussian probability density function,
\begin{equation}
\label{eq-wi}
w_k = \frac{1}{\sqrt{2\pi}}~ \int _{~~f_k-\frac{1}{2}\Delta f}^{f_k+\frac{1}{2}\Delta f}
\exp (-z^2/2)~ dz  ~~.
\end{equation}
\begin{figure}[t]
\begin{center}
\includegraphics[height=10cm,angle=0]{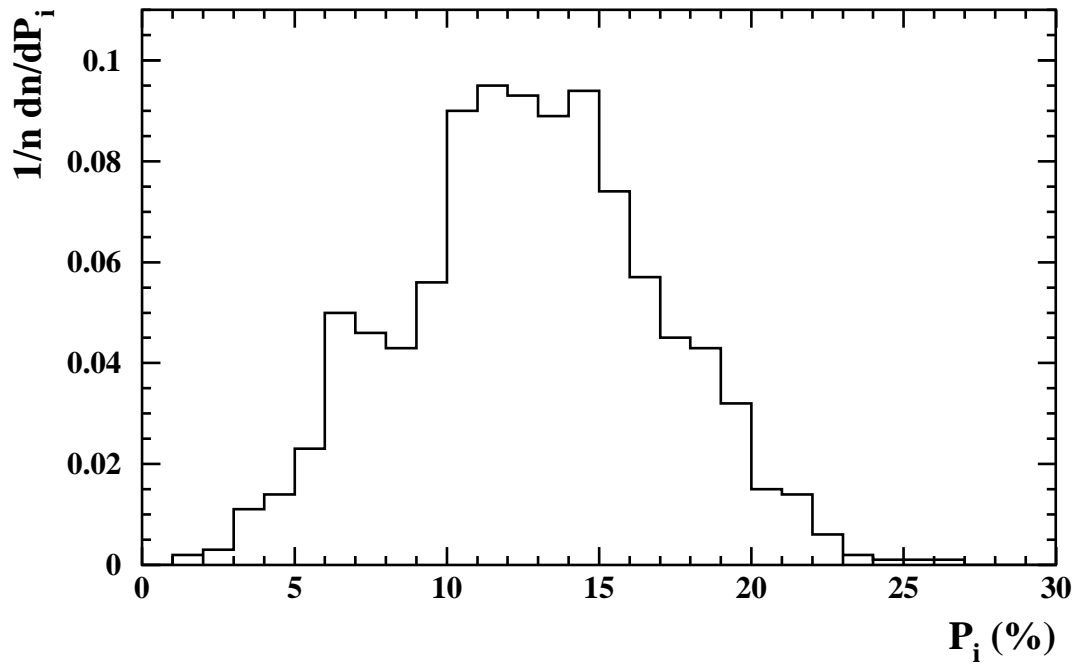}
\caption{Distribution of probabilities $P_i$ for photon primaries,
calculated as described in the text.}
\label{fig-photprob}
\end{center}
\end{figure}

The resulting distribution of $P_i$ for the photon profiles is plotted
in Figure~\ref{fig-photprob}.
Most profiles reach values well above 5\%.
Following Eq.~(\ref{eq-ptot}), an average probability $P \simeq 13\%$ is obtained.
This corresponds to a discrepancy between photons and data of about
1.5$\sigma$.
The results obtained for the primary photon and hadron hypotheses
are summarized in Table~\ref{tab-ptot}.

\begin{table}[t]
\begin{center}
\caption{Probability $P$ of a given primary particle hypothesis to be
consistent with the observed Fly's Eye event profile and corresponding
discrepancy $\Delta$ in units of standard deviations.}
\label{tab-ptot}
\vskip 0.5 cm
\begin{tabular}{lccc}
\hline
&        &  QGSJET~01  &  SIBYLL~2.1 \\
         & photon~~~ & p~~~~~C~~~~~Fe~ & p~~~~~C~~~~~Fe \\
$P$ [\%] & 13     & ~43~~~~54~~~~53~~ & 31~~~~52~~~~54
\\
$\Delta$ [$\sigma$] & 1.5 & 0.8~~~0.6~~~0.6 & 1.0~~~0.6~~~0.6
\\
\hline
\\
\end{tabular}
\end{center}
\end{table}
\section{Discussion}
\label{sec-conclusion}
For any considered primary hadron assumption and for both hadronic
interaction models, the values given in Table~\ref{tab-ptot}
confirm that no hypothesis can be rejected, as expected from the
$X_{max}$ investigation and in agreement with previous
considerations in the literature~\cite{event,halzen}.
Also in the framework of the SIBYLL~2.1 model, primary protons,
which on average reach the maximum $\simeq$70~g/cm$^2$ deeper in the
atmosphere than the observed profile, are not excluded due to the
measurement uncertainties and shower fluctuations.
They are only slightly less favoured (discrepancy of 1.0$\sigma$ to data)
with respect to heavier nuclei (0.6$\sigma$)
or primary protons in the QGSJET~01 model (0.8$\sigma$).

Compared to primary hadrons, photons are less favoured. With a discrepancy
to the Fly's Eye event profile on the level of 1.5$\sigma$, however,
also the primary photons are not excluded by the data.

This result does not confirm a previous analysis that claimed the primary
photon hypothesis to be inconsistent with the experimental 
observations~\cite{halzen}. Compared to our calculations, the 
shower maximum of primary photons had been obtained at depths larger by
$\simeq$90~g/cm$^2$.
Part of this difference might arise from the preshower
calculation, as might be expected due to the numerical error in
Ref.~\cite{erber} pointed out in Ref.~\cite{preshcors}.
Our analysis yields an average energy fraction carried
by particles with energies above $10^{19}$~eV of $\simeq 40\%$
(compare Figure~\ref{fig-espek}). As this is smaller than the value $>60\%$
given in~\cite{halzen}, a ``faster'' shower development is
expected, also due to the LPM effect being less important
for smaller particle energies.

The discrepancy between photons and data might be diminished below
1.5$\sigma$ by additional uncertainties not yet considered in the
quantitative treatment.
For the simulated profiles,
no ``theoretical'' uncertainty has been assumed when
testing the primary photon hypothesis.
Though much less affected by high-energy extrapolations of hadronic
interactions than hadron primaries,
uncertainties are associated, for instance, with the
photonuclear cross-section at the highest energies.
Given the specific conditions of the Fly's Eye event, photonuclear
reactions seem of minor importance.
For photons around $10^{19}$~eV, the expected photonuclear cross-section
\cite{block} is more than two orders of magnitude below the
pair production cross-section.
As preshowering additionally distributes
the initial primary energy among many particles entering the atmosphere,
photonuclear interactions of a small subset of these particles (of order
1 out of 300 particles) impose only a marginal effect on the overall
shower profile for the simulated primary photons.
Allowing, however, for a larger increase of the photonuclear cross-section
with energy, photon-initiated events would become more hadron-like.
The expected photon profiles would be shifted towards the observed one
which would reduce the discrepancy.

For the reconstructed data,
recent studies of the atmospheric density profile~\cite{bianca}
indicate a significant additional uncertainty in the reconstructed
atmospheric depth.
The air density profile is required for converting the {\it altitude}
of a shower track observed by the telescope to its derived {\it atmospheric
depth}, as the physics interpretation is mostly based on the latter quantity.
Without proper correction, variations of the density profile might lead
to misinterpretations of the reconstructed atmospheric depth of
10-20~g/cm$^2$ or more, depending on the specific conditions of the
observation.
Such density profile variations have been confirmed in investigations
of atmospheric data taken at Salt Lake City airport, i.e.~not far from
the Fly's Eye site,
that are available starting from
1996~\cite{barbara1}.
Moreover, a study of October density profiles measured over several years
at Salt Lake City airport indicates a systematic underestimation of about 
10$-$15~g/cm$^2$ on average in $X_{max}$ when using
the US standard atmosphere for geometries similar to the Fly's Eye
event~\cite{barbara2}.
Assuming such a shift in atmospheric depth of the reconstructed profile,
the discrepancy between photons and data would be
reduced by $\simeq$0.2$\sigma$.

Even without assuming a re-analysis of the Fly's Eye event that would
push the data closer to the photons, it seems fair to conclude that
additional uncertainties exist so that the potential to exclude a
specific primary particle type hypothesis is reduced. 
This strengthens the
conclusion that photons are not ruled out as primary particle for
the Fly's Eye event.

Clearly, a larger statistics of high-energy events is necessary to allow
stringent constraints on the photon hypothesis.
Let us illustrate the sensitivity of the Pierre Auger
Observatory~\cite{auger,augerea}
for giving an upper limit on the photon fraction in case 
no photon detection can be claimed.
Depending on the actual high-energy particle flux and including a
duty cycle of 10$-$15\%,
the fluorescence telescopes of the Auger experiment
are expected to record about 30$-$50 shower profiles with primary energies exceeding
$10^{20}$~eV within a few years of data taking.
For simplicity, let us consider a constant probability $\epsilon$
for each of these $n$ observed profiles to be originated by a photon.
Then, the probability that $n_\gamma$ out of $n$ profiles were photon-initiated is
\begin{equation}
\label{eq-prob}
P(n_\gamma) = \epsilon^{n_\gamma} ~ (1-\epsilon)^{n-n_\gamma} ~ (^{~~n}_{n-n_\gamma}) ~~~.
\end{equation}
For a choice of $n = 40$ and $\epsilon = 5\%$, with 95\% confidence level
a photon fraction exceeding 10\% could be excluded,
which would provide an important constraint for explaining the EHECR origin.
In case of the Fly's Eye event, a value of $\epsilon = 5\%$ corresponds
to reducing e.g.~the uncertainties $dX_j$ by a factor 1.5, which seems
well in reach for the Auger experiment. The sensitivity level from the
profile measurements alone might even be increased, as the most energetic
events will be observed from different telescope positions.
Further improvement might be gained by including data from the
Auger surface detectors. Apart from the virtually 100\% duty cycle of this
surface array, information on the muon content of the shower would
additionally strongly constrain
potential primary photons. The average muon number at observation level
for the events simulated in the current analysis is given in Table~\ref{tab-muon}.
In case of photon primaries, the muon number is reduced by more than 
a factor~3 relative to hadron primaries.
It is also interesting to note that the muon numbers predicted by 
QGSJET~01 and SIBYLL~2.1 differ quite significantly from each other. 
Compared to the overall profile,
the muon number is much more sensitive to different extra\-polations of hadronic
interaction features to the highest energies.
Combining the various observables
will help to improve the modeling of hadron-initiated showers.
\begin{table}[t]
\begin{center}
\caption{Simulated average muon number and RMS on observation level for the Fly's Eye event
conditions.}
\label{tab-muon}
\vskip 0.5 cm
\begin{tabular}{lccc}
\hline
&        &  ~~~QGSJET~01~~~  &  ~~~SIBYLL~2.1~~~ \\
& photon & p~~~~~~C~~~~~~Fe & p~~~~~~C~~~~~~Fe \\
  $N_{\mu} / 10^8$ & 3.1 & 12.9~~ 14.0~~ 15.3
            & 9.6~~ 11.0~~ 11.8
\\
  RMS($N_{\mu} / 10^8$) & 0.3 & ~1.5~~~~0.8~~~~ 0.6
            & ~1.2~~~~0.6~~~~0.5
\\
\hline
\\
\end{tabular}
\end{center}
\end{table}
{\it Acknowledgements.} We are grateful to Sergej Ostapchenko for drawing our
attention to the profile analysis.
We are indebted to Ralph Engel and Paul Sommers for comments on various
aspects of the investigation.
Very helpful discussions with Klaus Eitel on statistical methods 
and technical assistance from Dieter Heck for the simulations are kindly acknowledged.
This work was partially supported by the Polish State Committee for
Scientific Research under grants no.~PBZ KBN 054/P03/2001 and 2P03B 11024
and by the International Bureau of the BMBF (Germany) under grant
no.~POL 99/013.
MR is supported by the Alexander von Humboldt Foundation.

\end{document}